\documentclass[twocolumn,a4paper,11pt]{article}
\usepackage{blindtext}
\usepackage{titlesec}
\usepackage[utf8]{inputenc}
\usepackage{times}
\usepackage{microtype}

\usepackage{lscape}

\usepackage{epstopdf}

\usepackage[numbers,sort&compress]{natbib}
\usepackage{subfigure}
\usepackage{multirow}
\usepackage{float}
\usepackage{soul}
\usepackage{xcolor}
\graphicspath{{./images/}}

\usepackage{amssymb}
\usepackage{amsmath}
\usepackage{mathtools}
\usepackage{amsthm}
\usepackage{bm}

\usepackage[titletoc]{appendix}

\providecommand{\keywords}[1]
{
  \small	
  \textbf{\textit{Keywords---}} #1
}

\bibpunct[, ]{[}{]}{,}{}{,}{,}
\renewcommand\bibfont{\fontsize{10}{12}\selectfont}

\usepackage{authblk}
\title{Gaussian process with derivative information for the analysis of the sunlight adverse effects on color of rock art paintings\\}
\author[1$\ast$]{Gabriel Riutort-Mayol}
\author[2]{Michael Riis Andersen}
\author[3]{Aki Vehtari}
\author[1]{Jos\'e Luis Lerma}
\affil[1]{Department of Cartographic Engineering, Geodesy, and Photogrammetry, Universitat Polit\`ecnica de Val\`encia, Spain}
\affil[2]{Department of Applied Mathematics and Computer Science, Technical University of Denmark}
\affil[3]{Helsinki Institute for Information Technology HIIT, Department of Computer Science, Aalto University, Finland}
\affil[$\ast$]{gabriuma@gmail.com}
\date{}                     
\begin{document}

\maketitle

\begin{abstract}
Microfading Spectrometry (MFS) is a method for assessing light sensitivity color (spectral) variations of cultural heritage objects. The MFS technique provides measurements of the surface under study, where each point of the surface gives rise to a time-series that represents potential spectral (color) changes due to sunlight exposition over time. Color fading is expected to be non-decreasing as a function of time and stabilize eventually. These properties can be expressed in terms of the partial derivatives of the functions. We propose a spatio-temporal model that takes this information into account by jointly modeling the spatio-temporal process and its derivative process using Gaussian processes (GPs). 
We fitted the proposed model to MFS data collected from the surface of prehistoric rock art paintings. A multivariate covariance function in a GP allows modeling trichromatic image color variables jointly with spatial distances and time points variables as inputs to evaluate the covariance structure of the data. We demonstrated that the colorimetric variables are useful for predicting the color fading time-series for new unobserved spatial locations. Furthermore, constraining the model using derivative sign observations for monotonicity was shown to be beneficial in terms of both predictive performance and application-specific interpretability. 
\end{abstract}


\section{Introduction}\label{sec_intro}

Prehistoric rock art paintings are exposed to environmental elements, which can accelerate their degradation, increasing the risk of losing such a valuable piece of information of past societies. 
Apart from and in addition to many other factors, exposure to sunlight can have adverse effects on these systems due to thermal and photochemical degradation of the historic materials, and changes in the spectral properties of the materials is one of its main effects which is mainly related to the physicochemical properties of the materials \citep{Diez-Herrero_2009,del_hoyo_2015}. In this study we focused on the study and documentation of the degree of color changing/fading of paintings, patinas and host rock which is crucial for the conservation of these systems \citep{Cassar_2001}. 

It is known that materials with higher light sensitivity usually experience a rapid color change during the early stages of exposure, followed by a slower rate after maximum fading has occurred, assuming total disappearance of the substance that produces the color (chromophore) at this second stage of the fading \citep{feller1986determination,giles_1965,giles_1968,johnston1984kinetics}. Thus, fading can not decrease with time and it is expected to stabilize in the long term. 
Materials can show different times to saturation depending on their physicochemical properties and concentration of chromophores.

The microfading spectrometry (MFS) technique is a method  for assessing the susceptibility of cultural heritage objects to light fading \citep{Ford_2011,Ford_2013, Columbia_2013,Tse_2010}. 
Each measured point of the surface under study gives rise to a time-series that represents potential color fading due to light exposition over time \citep{Whitmore_1999, del_hoyo_2010}. Thus, MFS measurements can be seen as observations of an underlying spatio-temporal stochastic process. MFS time-series represents potential color fading  from the current state of the materials. 

The MFS instrument is very sensitive to movement and glossy surface effects, causing extremely large fluctuations in color fading values sometimes registered during measurements. Furthermore, they can be easily contaminated by changes in the illumination conditions when performed in outside environments, as it is the case of the surface of rock art paintings. These large fluctuations and possible systematic noise effects in the observations can cause that models do not satisfy those properties of monotonicity and long-term stabilization of color fading over time. So, it is recommendable to include additional information in the models in order to meet these properties and to ensure reliable properties for color fading estimates in new unobserved locations.

Existing lightfastness studies on these systems have been limited to analyze the few measured points on the surface of the rock art paintings due to the difficulty to set up the instrument, specially under harsh conditions.

So, in this paper we propose a reliable modeling framework, based on Gaussian processes (GPs) and their derivatives, aiming of rigorously extending the analysis of MFS color-fading in many others unobserved points on the surface of rock art paintings. Forecasting potential color-fading in every point of the surface of rock art paintings will be an important and useful information in order to achieve further successful conservation actions on these systems.

\subsubsection*{Background methods}
Functional data usually refers to independent realizations of a functional random variable that takes values in a continuous space (e.g. color-fading time-series, spatial or spatio-temporal observations).

In order to construct a model useful for making predictions of new functional data as function of new values of the variables in the input space, the process must be considered as an structured process with dependence among observations. 

Correlated functional models consider the observed functional data as non-independent functions. A popular approach for correlated functional data consists in penalized splines models \citep{wood2003thin} with different basis constructions based on Kronecker products \citep{currie2006generalized,lee2011p} or additive basis components \citep{kneib2006structured}. 

A geostatistical approach for spatio-temporal data is the kriging approach for 1D-functional data \citep{Giraldo_2010}, which the 1D-functional data consists in the time-series of the data. The spatial correlation of the time-series is modeled in the covariance function. This approach has the drawback of defining the same spatial structure for the whole time-series. A related approach can be found in \cite{Baladandayuthapani_2008} where the spatial correlation between the time-series is modeled by defining GPs with a spatial covariance function across the time-series functional coefficients. This construction allows for modeling different spatial structure for the different orders of the coefficients.

Another and powerful approach consists in considering the space-time structured observations as stochastic realizations of a GP prior with a spatio-temporal covariance function. GP \citep{Neal_1999, Rasmussen_2006} is a natural and flexible non-parametric prior model for multi-dimensional functions and with multivariate predictors (input variables) in each dimension. Furthermore, GP is sufficiently flexible to model complex phenomema since it allows possible non-linear effects and can handle interactions between input variables implicitly. 
GPs are easy to generalize/change to new model by changing the covariance function.
For a review of the different covariance functions in Gaussian processes, see \cite{Rasmussen_2006}. In a separable form, the space-time covariance function is a result of the interaction of the two independent processes, space and time \citep{banerjee_2014}. In a non-separable form, the covariance function models space-time interaction \citep{cressie1999classes,de2002nonseparable}. 

Several methods has been proposed for monotonic regression in the literature. The predominant focus of the theoretical literature on monotone function estimation has been on the methodology of order-restricted inference, which is sometimes also known as isotonic regression \citep{barlow1972statistical}. 
\cite{neelon2004bayesian} approach isotonic regression and order-restricted inference for non-parametric models in a Bayesian analysis. \cite{brezger_2008} induces monotonicity on penalized B-splines imposing order- restriction by specifying truncated prior distributions in order to reject the undesired draws for the parameters in the MCMC sampling. \cite{Reich_2011} makes a similar approach imposing order to the regression parameters by means of reparameterizing and constraining these parameters with application to a quantile regression model. \cite{shively_2009} proposes two approches to obtain monotonic functions, the first using a modified characterization of the smooth monotone functions proposed in \citep{ramsay_1998} that allows for unconstrained estimation, and the second using constrained prior distributions for the regression coefficients to ensure the monotonicity. 

In addition to monotonic regression imposed by construction, monotonicity can be expressed in terms of the sign of the partial derivative of the functions \citep{Solak_2003, Riihimaki_2011}. Furthermore, some constraints, such as saturation of the functions, can also be expressed in terms of the value of the partial derivatives. Linearity of parametric or semiparametric models makes feasible the use of their derivatives as additional observations jointly with the regular observations in order to force the function to fit these properties, e.g. monotonicity and saturation \citep{Rasmussen_2003}. In the same way, the derivative of a (mean-square differentiable) GP function remains a GP because differentation is a linear operator \citep{Solak_2003}. This makes it possible to use derivative observations jointly with regular observations in a Gaussian process model, by extending the covariance function accordingly to include the covariances between the process and its partial derivatives \citep{Solak_2003}. \cite{Riihimaki_2011} proposed a method for imposing monotonicity onto GP by introducing virtual observations of the sign of the derivative of the process at a discrete set of locations in the input space, which can be interpreted as a probabilistic version of the idea of monotonicity hints \citep{sill1998monotonic}. The same mechanism is used to induce monotonicity in the deep probabilistic models \citep{lorenzi2018constraining}. It is known that there are practical issues with this approach of inducing monotonicty through virtual observations, since the posterior distribution of the functions depends on the number of the inducing points. When this is too large, the posteriors tend to be overly smoothed. However, if the function is quite smooth, this problem can be avoided in practice choosing few inducing points only.

\subsubsection*{Objectives and methodology of the paper}
In this paper, a specific application with the aim of modeling and predicting MFS color fading time-series for new unobserved spatial locations on the surface of rock art paintings is presented. 
The main motivation of this study is to construct a model that exploits to the maximum the correlation structure of the data in order to extend the analysis and make useful predictions, in an scenario where a small set of sampling observations are available, as it is the case of MFS on rock art paintings; In fact, in the current study only 13 measured locations on the surface were provided. A multivariate GP prior model is the more natural way to accomplish this objective for this type of data. 

A multi-dimensional (e.g. spatial and temporal) covariance function is the key element of a GP as it encodes the functional relationship and defines the correlation structure which characterize the correlation between function values at different inputs. Furthermore, the covariance function in a GP can be extended to jointly model the covariances of regular and derivatives observations, increasing thus the predictive capacity of the model and, at the same time, guaranteeing that the predictions are monotonically non-decreasing and stabilize eventually as a function of time.

We propose a spatio-temporal model that takes the derivative information into account by jointly modeling the regular process and its derivative process using GPs. Derivative observations of both the sign and the values of partial derivatives are used to induce monotonicity (non-decreasing) and long-term saturation as a function of time. Furthermore, in order to force the functions to be zero at the starting time points ($t=0$), noise-free pseudo-observations are used at these points.

Physicochemical data for all the points on the surface to predict new curves are hard to obtain. Instead, image color values can be used as input variables to construct and evaluate the correlation, since these image color variables are related to the physicochemical properties of the imaged surface \citep{Malacara_2011}. A multivariate covariance function in a GP allows modeling trichromatic image color variables jointly with spatial distances and time points variables as inputs to evaluate the covariance structure of the data.

In order to do model evaluation and comparison, the same model but without derivative information is also fitted. Cross-validation procedures are conducted to compute the \textit{posterior predictive checks}, the \textit{expected log predictive density} and the \textit{mean square predictive errors} in order to do model checking and assessment of the predictive performance.

The rest of the paper is structured as follows. Section \ref{sec:model} focuses on the modeling and inference formulation and Section \ref{sec_check} on the model checking and model selection procedures. Section \ref{sec:data} describes the case of study and the available data in detail. Section \ref{sec:results} describes the results of applying the proposed model on the data set. Section \ref{discussion} discusses about the results and modeling. Finally, Section \ref{conclusion} presents a brief conclusion of the work.

\section{Gaussian process model with derivative observations\label{sec:model}}

\subsubsection*{Gaussian process model}
We consider a continuous stochastic process based on a collection of spatio-temporal output observations $\bm{y} \in {\rm I\!R}^{NT}$ with associated inputs $X\in {\rm I\!R}^{NT\times D}$, where $y_{it} \in {\rm I\!R}$ denotes the value of the $i$th location at time $t$, with $i=1,\dots,N$ representing the spatial locations and $t=1,\dots,T$ the time points, and $D$ is the number of inputs variables. We adopt the model $\bm{y}=\bm{f} + \bm{e}$, where $\bm{f}$ is modeled as a Gaussian process and $\bm{e}$ is a Gaussian noise term. So, the observational model for the data can be written as follows
\[
p(\bm{y}_{it}|\bm{f}_{it})  =  \prod_{i,t}\mathtt{N}(y_{it}|f_{it},\sigma^2)
\]

\noindent where $f_{it}$ is the value of the function $f$ evaluated at $X_{it}$, i.e. $f_{it} = f(x_{it})$, and $\sigma^2$ is the noise variance. A zero mean Gaussian process prior with covariance matrix $K$ is assumed for the vector $\bm{f} \in {\rm I\!R}^{NT}$ of latent values, $p(\bm{f}|X) = \mathtt{N}(\bm{f} | 0,K(X;\theta))$. The covariance matrix $K \in {\rm I\!R}^{NT \times NT}$ is the result of the Kronecker product, $K = K_S(X_S; \theta_S) \otimes K_T(X_T; \theta_T)$, between the spatial covariance matrix $K_S \in {\rm I\!R}^{N \times N}$ and time covariance matrix $K_T \in {\rm I\!R}^{T \times T}$. With $X_S \in {\rm I\!R}^{N \times D_S}$ containing the inputs in the spatial dimension and $X_T \in {\rm I\!R}^{T \times D_T}$ containing the inputs in the temporal dimension, and $X=X_S \otimes X_T$ with $D=D_S+D_T$.  For both, the spatial and temporal covariance matrices, we use exponentiated quadratic covariance functions of vectors of parameters $\theta_S$ and $\theta_T$, respectively. The resulting covariance function $K$ from this Kronecker product takes the form
\begin{eqnarray*}
\lefteqn{ K(X;\theta)_{(i,t),(j,k)} = }    \\
&&\alpha^2 \exp\left( -\frac{1}{2} \sum_{d=1}^{D}\frac{1}{\rho^2_d}(x_{(i,t),d}-x_{(j,k),d})^2\right), 
\end{eqnarray*}

\noindent where $i,j=1,\dots,N$ and $t,k=1,\dots,T$, and $\theta$ contains the parameters $\alpha$ and $\rho=\{\rho_d\}_{d=1}^D$. The hyperparameter $\alpha$ is the prior standard deviation of the latent Gaussian process $\bm{f}$. The lengthscale hyperparameter $\rho_d$ control the smoothness of the function in the direction of the $d$th predictor (input variable). A exponentiated quadratic function assumes stationarity with respect to the input variables, and using a Kronecker product implies separability with respect to the spatial and temporal input dimensions. 

\subsubsection*{Function value constraints (zero-order constraints)}
In order to ensure that the function values are zero at the starting time points, that is, the time-series must start in zero, $f(X_A)=0$, where $A=\{(i,t):t=0\}$ is the subset of starting time points.
This property can be specified by using virtual observations equal to zero at these points, $\{y_A=0\}$, and the Dirac Delta function $\delta$ as an observational model for these observations,
\begin{equation}\label{eq:regularnoisefree}
p(\bm{y}_{A}|\bm{f}_{A})  = \delta(\bm{y}_{A} - \bm{f}_{A}).
\end{equation}

\noindent While the rest of observations, $\{(i,t): t\neq 0\}\notin A$, are considered to be contaminated with Gaussian noise
\begin{equation}\label{eq:regularnoise}
p(\bm{y}_{-A}|\bm{f}_{-A},\sigma^2)  =  \prod_{i,t\not\in A}\mathcal{N}(y_{it}|f_{it},\sigma^2),
\end{equation}

\noindent where $\bm{y}_{-A}$ denote the dataset $\bm{y}$ without the subset $A$ of observations, $\bm{y}_{-A}=\{y_{it}:(i,t)\notin A \}$. And $\bm{f}_{-A}$ denotes the latent function values $\bm{f}$ without the subset $A$ of observations, $\bm{f}_{-A}=\{f_{it}:(i,t)\notin A \}$.

\subsubsection*{Gaussian process with derivatives}
Apart from the regular observations $\bm{y}$, we want to include partial derivative observations and first order constraints in the model. Color degradation is expected to be non-decreasing as a function of time and stabilize in the long term. These properties can be expressed in terms of the first order partial derivative of the functions. It is possible to jointly model a function and its derivatives using a mean-square differentiable Gaussian process model, by extending the covariance function accordingly to include the covariances between the process and its partial derivatives. Hence, we can write the joint prior distribution for the latent function values $\bm{f}$ and the values of the partial derivatives of the latent function $\bm{f}'$ as follows
\begin{flalign} \label{eq:gpprior}
& p(\bm{f},\bm{f}'|X,X',\theta) =   \nonumber \\ 
& \mathcal{N} \left( \left[ \begin{array}{cc}
\bm{f} \\ 
\bm{f}'
\end{array} \right] | 0,\left[ \begin{array}{cc}
K_{f,f}(X,\theta) & K_{f,f'}(X,X',\theta) \\ 
K_{f',f}(X,X',\theta) & K_{f',f'}(X',\theta)
\end{array} \right] \right),
\end{flalign}

\noindent where $\bm{f}'$ denotes the values of the partial derivatives of $\bm{f}$ with respect to some input dimension evaluated at $X'$. The covariance matrix is extended to include the covariances between observations and partial derivatives ($K_{f,f'}$ and $K_{f',f}$) and the covariances between partial derivatives ($K_{f',f'}$). The covariance between a partial derivative and a function value is given by
\begin{eqnarray}
\lefteqn{ {\mathrm Cov}\left[\frac{\partial f^{(i,t)}}{\partial x^{(i,t)}_g},f^{(j,k)} \right] = }  \nonumber \\ 
&& \alpha^2 \, \exp\left( -\frac{1}{2} \, \sum_{d=1}^{D}\rho^{-2}_d(x_{(i,t),d}-x_{(j,k),d})^2\right)   \nonumber \\
&&  \times~ \left( -\rho^{-2}_g(x_{(i,t),g}-x_{(j,k),g})\right), 
\end{eqnarray}

\noindent and the covariance between partial derivatives
	\vspace{0cm}
\begin{flalign}
& {\mathrm Cov}\left[\frac{\partial f^{(i,t)}}{\partial x^{(i,t)}_g},\frac{\partial f^{(j,k)}}{\partial x^{(j,k)}_h} \right] =  \nonumber \\
&\alpha^2 \, \exp\left( -\frac{1}{2} \, \sum_{d=1}^{D}\rho^{-2}_d(x_{(i,t),d}-x_{(j,k),d})^2\right) \nonumber \\
& \times \rho^{-2}_g \left( \delta_{gh}-\rho^{-2}_h(x_{(i,t),h}-x_{(j,k),h})(x_{(i,t),g}-x_{(j,k),g})\right), 
\end{flalign}

\noindent In the previous equation, $\delta$ is the Kronecker Delta function where $\delta_{gh}=1$ if $g=h$, and 0 otherwise \citep{Riihimaki_2011}. We can now combine the joint prior distribution in eq. (\ref{eq:gpprior}) with an observation model for the partial derivatives, $p(\bm{y}'|\bm{f}')$, to encode information about the partial derivatives of $\bm{f}$ into the model. In this work, we consider two types of derivative observations: observations of the value of a partial derivative $y'_{it} \in {\rm I\!R}$ and observations of the sign of a partial derivative $z'_{it}={\mathrm sign}(f'_{it}) \in [1,-1]$, where 1 means that the partial derivative of the function is positive at the given data point and $-1$ means that the partial derivative is negative (decreasing function). 

\subsubsection*{Derivative constraints (first-order constraints)}
In order to impose a saturation constraint for long term stabilization, at the subset of ending time points $B=\{(i,t): t=T\}$, virtual observations of the values of partial derivatives with respect to the time input variable equal to zero, $\{y_{it}' = \frac{\partial f^{(i,t)}}{\partial t^{(i,t)}} = 0: (i,t)\in B\}$, are considered. The Dirac Delta function $\delta$ is used as observational model for these observations,
\begin{equation}\label{eq:derivativenoisefree}
p(\bm{y}'_{B}|\bm{f}'_{B})  = \delta(\bm{y}'_{B} - \bm{f}'_{B}).
\end{equation}

Furthermore, the function is guaranteed to be non-decreasing as a function of time when the partial derivative is non-negative. So, virtual observations of the sign of the partial derivatives with respect to the time input variable equal to one, $\{z_{it}= {\mathrm sign} \big( \frac{\partial f^{(i,t)}}{\partial t^{(i,t)}} \big) = 1; (i,t)\in C\}$, can be considered, where $C$ is the subset of desired time points where to induce monotonicity. The probit function $\Phi: {\rm I\!R} \to (0,1)$ can be used as a likelihood for the signs of the partial derivatives,
\begin{equation}\label{eq:probit}
p(\bm{z}_{C}|\bm{f}'_{C}) =  \prod_{i,t\in C}\Phi\left(z_{it} \cdot f'_{it} \cdot \frac{1}{v} \right).
\end{equation}

\noindent The function $\Phi$ is the standard Normal cumulative distribution function  and $v > 0$ is a parameter controlling the strictness of the constraint. When $v$ approaches zero ($v \to 0$), the function $\Phi$ approaches a step function. 

\subsubsection*{Likelihood function}
The joint likelihood of both regular $\bm{y}$ and derivative $\bm{y}'$ observations, given latent functions $\bm{f}$ and $\bm{f}'$, and hyperparameter $\sigma$, results:
\begin{eqnarray} \label{eq:likeli}
\lefteqn{ p(\bm{y},\bm{y}'|\bm{f},\bm{f}',\sigma) = } \nonumber \\
&& \prod_{i,t\not\in A}\mathcal{N}(y_{it}|f_{it},\sigma^2) \times \delta(\bm{y}_{A} - \bm{f}_{A}) \nonumber \\
&& \times \prod_{i,t\in C}\Phi\left(z_{it} \cdot f'_{it} \cdot \frac{1}{v} \right) \times \delta(\bm{y}'_{B} - \bm{f}'_{B})
\end{eqnarray}

\subsubsection*{Bayesian inference}
Bayesian inference is based on the posterior joint distribution of parameters given the data, which is proportional to the product of the likelihood and prior distributions,
\begin{eqnarray*}
\lefteqn{ p(\bm{f},\bm{f}',\sigma|\bm{y},\bm{y}') \propto } \\
&& p(\bm{y},\bm{y}'|\bm{f},\bm{f}',\sigma) p(\bm{f},\bm{f}'|X,X^{*},\theta)p(\theta)p(\sigma),
\end{eqnarray*}

\noindent where $p(\bm{y},\bm{y}'|\bm{f},\bm{f}',\sigma)$ is the likelihood of the model in eq. (\ref{eq:likeli}) and $p(\bm{f},\bm{f}'|X,X^{*},\theta)$ is the joint Gaussian process prior of regular $\bm{f}$ and derivative $\bm{f}'$ latent functions in eq. (\ref{eq:gpprior}). We set positive half-normal prior distributions for the hyperparameters $\alpha$, $p(\alpha)=\mathcal{N}^+(\alpha|0,1)$, and $\sigma$, $p(\sigma)=\mathcal{N}^+(\sigma|0,1)$, and gamma distributions for the hyperparameters $\bm{\rho}$, $p(\rho_d)={\mathrm Gamma}(\rho_d|1,0.1)$ for all $d$. These correspond to weakly informative prior distributions based on prior knowledge about the magnitude of the parameters. Thus, the joint posterior distribution can be expressed as depicted in eq. (\ref{eq_joint_posterior}).

\begin{figure*}[!t]
\begin{eqnarray} \label{eq_joint_posterior}
\lefteqn{ p(\bm{f},\bm{f}',\sigma|\bm{y},\bm{y}') \propto } \nonumber\\
&& \prod_{i,t\not\in A}\mathcal{N}(y_{it}|f_{it},\sigma^2) \times \delta(\bm{y}_{A} - \bm{f}_{A}) \nonumber \\
&& \times \prod_{i,t\in C}\Phi\left(z_{it} \cdot f'_{it} \cdot \frac{1}{v} \right) \times \delta(\bm{y}'_{B} - \bm{f}'_{B})  \nonumber \\
&& \times \prod_{i,t} \mathcal{N} \left( \left[ \begin{array}{cc}
f_{it} \nonumber \\ 
f'_{it}
\end{array} \right] | 0,\left[ \begin{array}{cc}
K_{f,f}(X,\theta)_{it} & K_{f,f'}(X,X^{*},\theta)_{it} \nonumber \\ 
K_{f',f}(X,X^{*},\theta)_{it} & K_{f',f'}(X^{*},\theta))_{it}
\end{array} \right] \right)  \nonumber \\
&& \times \mathcal{N}^+(\alpha|0,1)   \prod_{d=1}^D {\mathrm Gamma}(\rho_d|1,0.1) \mathcal{N}^+(\sigma|0,1). 
\end{eqnarray}
\end{figure*}

Predictive inference for new function values $\bm{\tilde{y}}$ for a new sequence of input values $\tilde{X}$ can be computed by integrating over the joint posterior distribution,
\begin{eqnarray}\label{eq_pred_posterior}
\lefteqn{ p(\bm{\tilde{y}},\bm{\tilde{y}'}|\bm{y},\bm{y}') } \nonumber\\
&\propto & \int p(\bm{\tilde{y}},\bm{\tilde{y}'}|\bm{\tilde{f}},\bm{\tilde{f}'},\sigma) p(\bm{\tilde{f}},\bm{\tilde{f}'}|\bm{f},\bm{f}',\theta) \nonumber \\
&&\cdotp(\bm{f},\bm{f}',\theta,\sigma|\bm{y},\bm{y}') d\bm{f} d\bm{f}' d\theta
d\sigma.
\end{eqnarray}

To posterior distribution of interest $p(\bm{f},\sigma|\bm{y},\bm{y}')$ is in general intractable. Hence, to estimate both parameter posterior distribution and posterior predictive distribution for this model, simulation methods and/or distributional approximations methods \citep{Gelman_2013} must be used. Simulating methods based on Markov chain Monte Carlo \citep{Brooks_2011} and, more recently, on Hamiltonian Monte Carlo \citep{neal_2011} are general sampling methods to obtain samples from the joint posterior distribution. For large data sets where iterative simulation algorithms are too slow, modal and distributional approximation methods can be efficient and approximate alternatives \citep{Gelman_2013}.

\section{Model checking, predictive performance and model selection}\label{sec_check}
Common procedures of checking normality and tendencies on the fitted residuals can be used in order to check the model. Additionally, more rigorous procedures ---in order to guarantee good model performance and ensure that the model is compatible with the data--- are the \textit{posterior predictive checks}, the \textit{expected log predictive density} and the \textit{mean square predictive errors}, which can be computed following cross-validation procedures. 

The \textit{posterior predictive checks}, which are also known as the \textit{leave-one-out probability integral transformation} (LOO-PIT), are based on computing the probability of new predictions to be lower or equal to its corresponding actual observed values \citep{gelfand_1992,Gelman_2013} following a leave-one-out cross-validation procedure,
\begin{equation*} \label{eq:ppc}
\text{LOO-PIT}_{(i,t)\in E} = p(\tilde{y}_{(i,t)\in E} \leq y_{(i,t)\in E}),
\end{equation*}

\noindent where $E$ is the subset of observation indices to leave out in a step of the cross-validation. $\tilde{y}_{(i,t)\in E}$ is the new observation from the predictive distribution, and $y_{(i,t)\in E}$ is the actual observation. The similarity or provenance of these probabilities from standard uniform distributions endorses these cross-validation probabilities with the desirable property of having the same interpretation across models, which implies good fit to the data \citep{Bayarri_2000}. Using simulation methods for estimating and predicting a Bayesian model, computing the probability of a predicted value to be lower than the observed one is straightforward through the collection of simulated values.

The \textit{expected log predictive density} (ELPD) evaluates, by averaging over all the steps in the cross-validation procedure, how far new data is from the model while taking the posterior uncertainties into account. It is based on the log-density of new data given the model \citep{vehtari_2012}:
\begin{equation*} \label{eq:lppd}
\text{ELPD}_{E} = \frac{1}{|E|} \sum_{(i,t)\in E} ln (p(y_{it}|\bm{y}_{-E})),
\end{equation*}

\noindent where $|E|$ denotes the cardinality of the subset $E$ of observation indices to leave out in a step of the cross-validation. $\bm{y}_{-E}$ denotes the dataset without the subset $E$ of observations.

Finally, the \textit{mean square error} (MSE) of the predictions, also evaluates, by averaging over all the steps in the cross-validation procedure, how far new data is from the model by using the distance (error) between the actual observation $y_{(i,t)\in E}$ and the predictive mean $\tilde{y}_{(i,t)\in E}|\bm{y}_{-E}$:
\begin{equation*} \label{eq:spe}
\text{MSE}_{E} = \frac{1}{|E|} \sum_{(i,j)\in E} (y_{it} - \tilde{y}_{it})^2.
\end{equation*}

Three different cross-validation scenarios are conducted to compute and evaluate these statistics in order to do model checking, assessment of predictive performance, and model comparison between the model with derivative information and the model without derivative information. 

\subsubsection*{\textit{\textmd{Leave one observation out cross-validation (CV1):}}}
\hspace{2mm} In this case, the subset $E$ of observation indices will be just a single observation $(i,t)$. The statistics LOO-PIT, ELPD, and MSE will be computed in this leave-one-out cross-validation scenario. The LOO-PIT is essentially useful for model checking, ensuring that the model is compatible with the data. The ELPD and the MSE evaluate the predictive performance of individual observations $(i,t)$.

\subsubsection*{\textit{\textmd{Leave one location out cross-validation (CV2):}}}
\hspace{2mm} The end goal of this work is to predict complete color-fading time-series at new unobserved locations. So, in this case the subset of observation indices $E$ will be a complete time-series of a specific spatial location $i$, $E=\{(i,t): t\in \{1,\dots,T\}\}$. The statistics ELPD and MSE will be computed in this cross-validation scenario. Plots of predicted new time-series superimposed to their corresponding actual observations will be shown in order to visually evaluate the predictive performance. Model selection can be done by comparing the predictive performance between models using the ELPD and MSE statistics. The best model is who maximizes the ELPD and/or minimizes the MSE.

\subsubsection*{\textit{\textmd{Leave the last part of a time-series out cross-validation (CV3):}}}
\hspace{2mm} Evaluating the model in terms of the extrapolation of the time-series can be of interest. In this case, the subset of observation indices $E$ will be the last part of a time-series of a specific spatial location $i$, 
\begin{equation}\label{ell}
E=\{(i,t): t\in \{T-\ell-1,\dots,T\}\},
\end{equation}
\noindent where $\ell$ denotes the slide of the time-series considered. The statistics ELPD and MSE will be computed in this cross-validation scenario. Plots of predicted extrapolated time points superimposed to their corresponding actual observations will be shown in order to visually evaluate the predictive performance. Model selection can be done by comparing the ELPD and MSE statistics.

\section{Case study and data description} \label{sec:data}
In this practical case we seek to evaluate the degree of color fading over time and space on rock art paintings caused by direct solar irradiation. The goal is to construct a model for modeling the set of sampling MFS data and making predictions at new unobserved locations under the surface of rock art paintings as a function of new input values. 

The study area is located in the cova Remigia rock art shelter, Castell\'on (Spain). Some of its paintings, which are included in UNESCO’s World Heritage List, are exposed to environmental factors, including the natural daylight depending on the time of the day and the season of the year. Some authors have demonstrated that exposure to sunlight which can have adverse effects on these systems due to thermal and photochemical degradation of the historic materials \citep{del_hoyo_2015}. 

\begin{figure}[H]
\centering
\subfigure{\includegraphics[width=0.5\textwidth, trim = 15mm 20mm 0mm 25mm, clip]{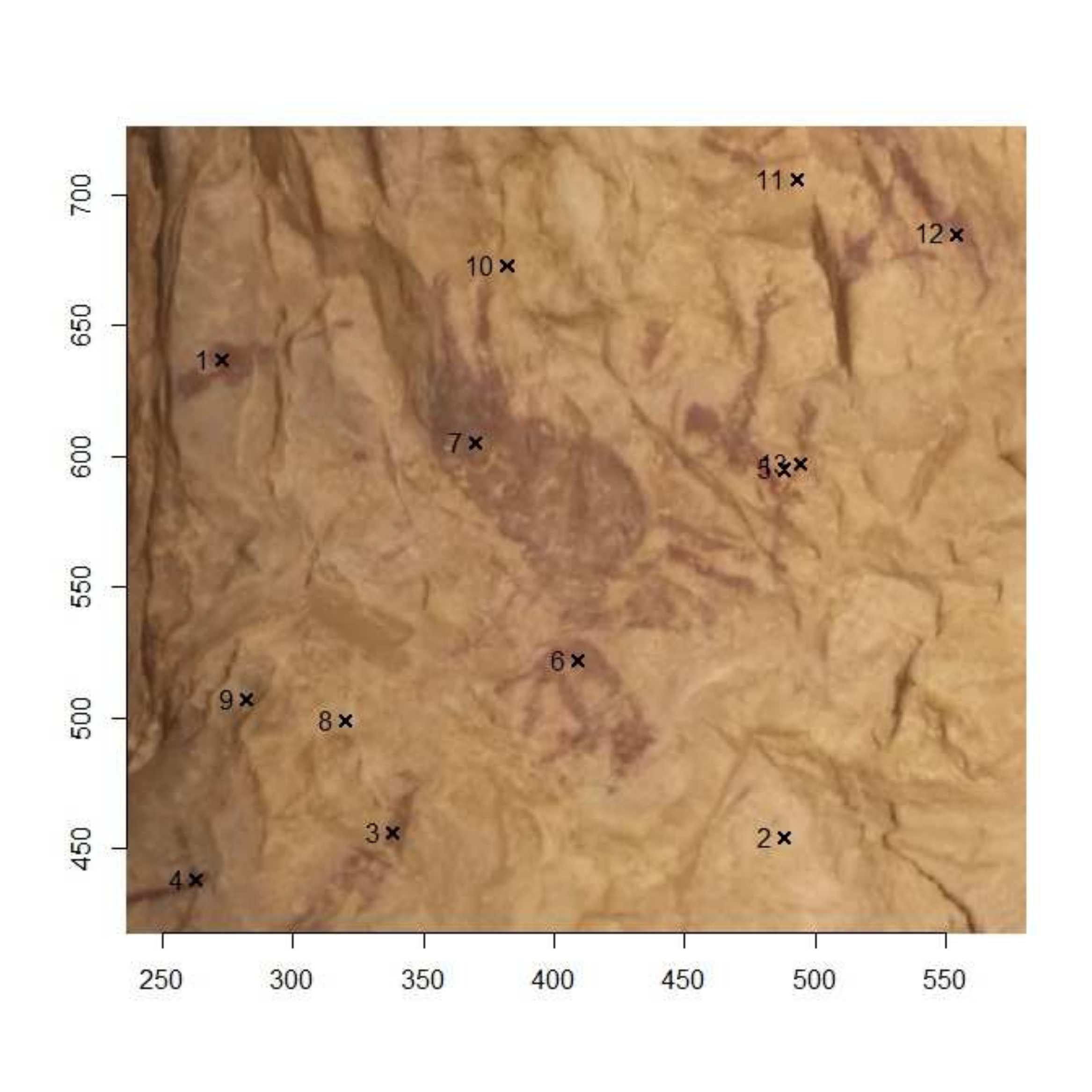}}
\caption{Rock art paintings image with the points where the MFS observations were measured. This plot is in pixel units and corresponds to a clip from a bigger image. 
}
  \label{image_observed}
\end{figure}

Each measured point on the surface gives rise to a time-series that represents color fading (Color differences/Delta E* value, \cite{Malacara_2011}) over time. The MFS measurements have a duration of 10 minutes. The sampling frequency will be once per minute, such that the resulting time-series will consist of 11 time points $(t=0,\dots,10)$. Thus,  the spatio-temporal MFS data set consist of 13 observed locations on the surface ($N=13$). Figure \ref{image_observed} shows their pixel locations on a color image of the study area. The observed time-series of values can be seen plotted as crosses in Figure \ref{fig:fitted}. The available input variables for every spatial location are the three image color variables, Hue ($H$), Saturation ($S$) and Intensity ($I$), and the two spatial coordinates $S_x$ and $S_y$. In the temporal dimension the input variable is the collection of time points, $t=0,...,10$, of the MFS time series. Table \ref{tab:summary} presents summary statistics for the input variables $H$, $S$, $I$, $S_x$, and $S_y$. These input variables have been re-scaled by dividing by their standard deviations. In the case of $S_x$ and $S_y$, they were divided by their common standard deviation.

Due to the large fluctuations present in these measurements conducted on these rock art systems, some modeling issues can arise, such as not starting at zero or not being monotonically increasing or not stabilizing in the long term as a function of time, which are properties assumed for the color fading curves, as discussed in the above. 

\begin{table}[H]
\caption{\label{tab:summary}Summary statistics of the input variables}

\centering
\fbox{
\begin{scriptsize}
\begin{tabular}{l|ccccc}
  & H & S & I & $S_x$ & $S_y$ \\
\hline
Mean & 5.255 & 9.704 & 5.155 & 3.549 & 4.969 \\
Std. Dev. & 1.0 & 1.0 & 1.0 & 0.732 & 0.674 \\
1st Qu. & 4.359 & 9.042 & 4.603 & 2.910 & 4.387 \\
3rd Qu. & 5.913 & 10.273 & 5.772 & 4.187 & 5.551 \\

\end{tabular}
\end{scriptsize}
}
\end{table}

The derivative process to induce monotonicity through virtual observations of the sign of the partial derivative, as explained in Section \ref{sec:model}, has only been observed at two time points of every location $\{(i,t):t\in\{6,9\}\}$, preventing in this way the inclusion of too many inducing points and to obtain too smooth posterior samples. On the other hand, the eventually saturation constraint of the color fading curves is not taken into account in the modelling of this practical case because 10 minutes of measurement does not seem long enough to consider fading having reached saturation. 

The value of the parameter $v$ of the probit likelihood in equation (\ref{eq:probit}) is set to $10^{-4}$. The slide $\ell$ in equation (\ref{ell}) in the cross-validation scheme CV3 is set to 7 time points.

A comparison with the same model but without derivative information is carried out and evaluated in terms of both predictive performance and application-specific interpretability.

The actual equivalency of the exposure time used in MFS in years depends on the hours and intensity of sunlight that affects the paintings on a changing everyday basis. Without proper monitoring of light, this equivalency is difficult to obtain, so this aspect of the research was not considered in the current study.

\section{Experimental results\label{sec:results}}

The posterior distributions and predictive distributions have been estimated by Hamiltonian Monte Carlo sampling methods \citep{neal_2011} using the Stan software \citep{Carpenter_2017}. Three simulation chains with different initial values have been launched. The convergence of the simulation chains was evaluated by the split-Rhat convergence diagnosis \citep{Gelman_2013} and the effective sample size of the chains. A value of 1 in the split-Rhat convergence statistic indicates convergence of the simulation chain, although conventionally accepted values of convergence would be between 1 and 1.1. In our case, a split-Rhat value lower than 1.05 has been obtained for all parameter simulation chains.

The input variables $H$, $S$ and $I$ were previously re-scaled by dividing by their standard deviations. The $S_x$ and $S_y$ spatial coordinates were divided by their common standard deviation. The input variable $t$ was not re-scaled. The estimated posterior distributions, $p(\theta|\bm{y},\bm{y}')$, for the hyperparameters $\theta$ of the model can be visualized in Figure \ref{param} in Appendix \ref{app_hist_post}. The hyperparameter $\alpha$ (with an estimated mode of 2.4) is the overall scale of the latent Gaussian process prior. The hyperparameter $\sigma$ (with an estimated mode of 0.37) is the standard deviation of the observations. The hyperparameter $\rho_d$ ($d=1,\dots,5$) is the lengthscale parameter of the exponentiated quadratic covariance function of the Gaussian process prior associated with the $d$th input variable. The spatial coordinates input variables $S_x$ and $S_y$ are sharing the lengthscale parameter $\rho_4$ (with a estimated mode of 4.5), such that the covariance function depends on the (Euclidean) distance between spatial coordinates. The hyperparameter $\rho_5$ (with a estimated mode of 8.2) is the lengthscale parameter associated to the time input variable, and the hyperparameters $\rho_1$, $\rho_2$, and $\rho_3$ (with estimated modes of 0.95, 18.3, and 1.3, respectively) are the lengthscale parameters associated to the $H$, $S$, and $I$ variables, respectively. The posterior distributions of the process, $p(\bm{f}|\bm{y},\bm{y}')$, versus the input variables $H$, $S$ and $I$, and the time points, can be visualized in Figure \ref{b-vectors} in Appendix \ref{app_post_vs_inputs}. 

Figure \ref{fig:fitted} shows the predictive distributions (predictive means and 95\% pointwise predictive credible intervals) of the regular process, $p(\tilde{\bm{y}}|\bm{y},\bm{y}')$, as a function of time for specific spatial locations. Additionally, the predictive means for the Gaussian process model without derivative information are also plotted for comparison. The predictive distributions of the derivative process, $p(\bm{f}'|\bm{y},\bm{y}')$, as a function of time are plotted for specific spatial locations in Figure \ref{fig:fitted_and_derivatives} in Appendix \ref{app_pred_derv} .

 \vspace{1cm}

\begin{figure*}[h]
\centering
\subfigure{\includegraphics[width=0.325\textwidth,trim = 0mm 0mm 165mm -5mm, clip]{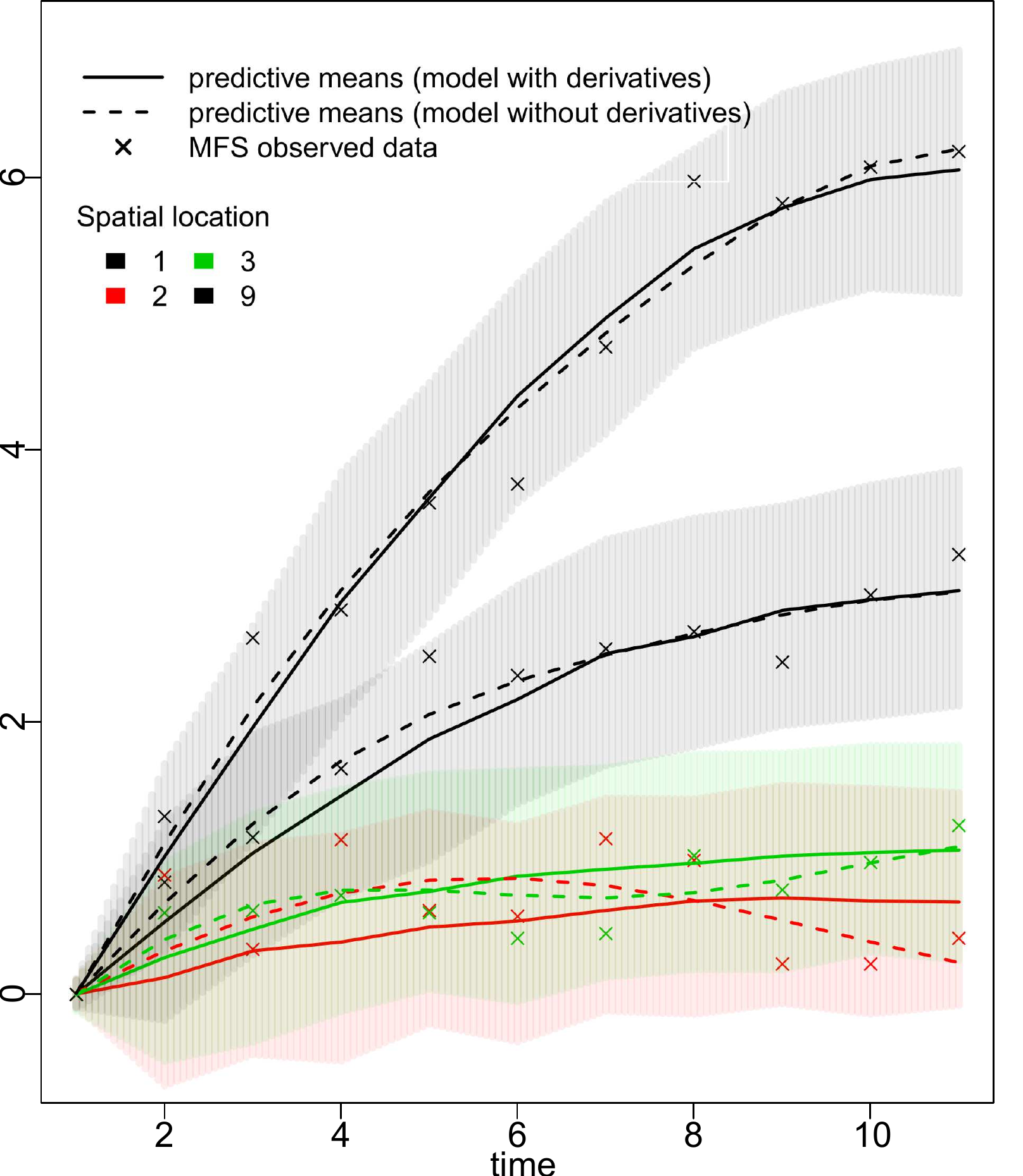}}
\subfigure{\includegraphics[width=0.325\textwidth,trim = 0mm 0mm 163mm -5mm, clip]{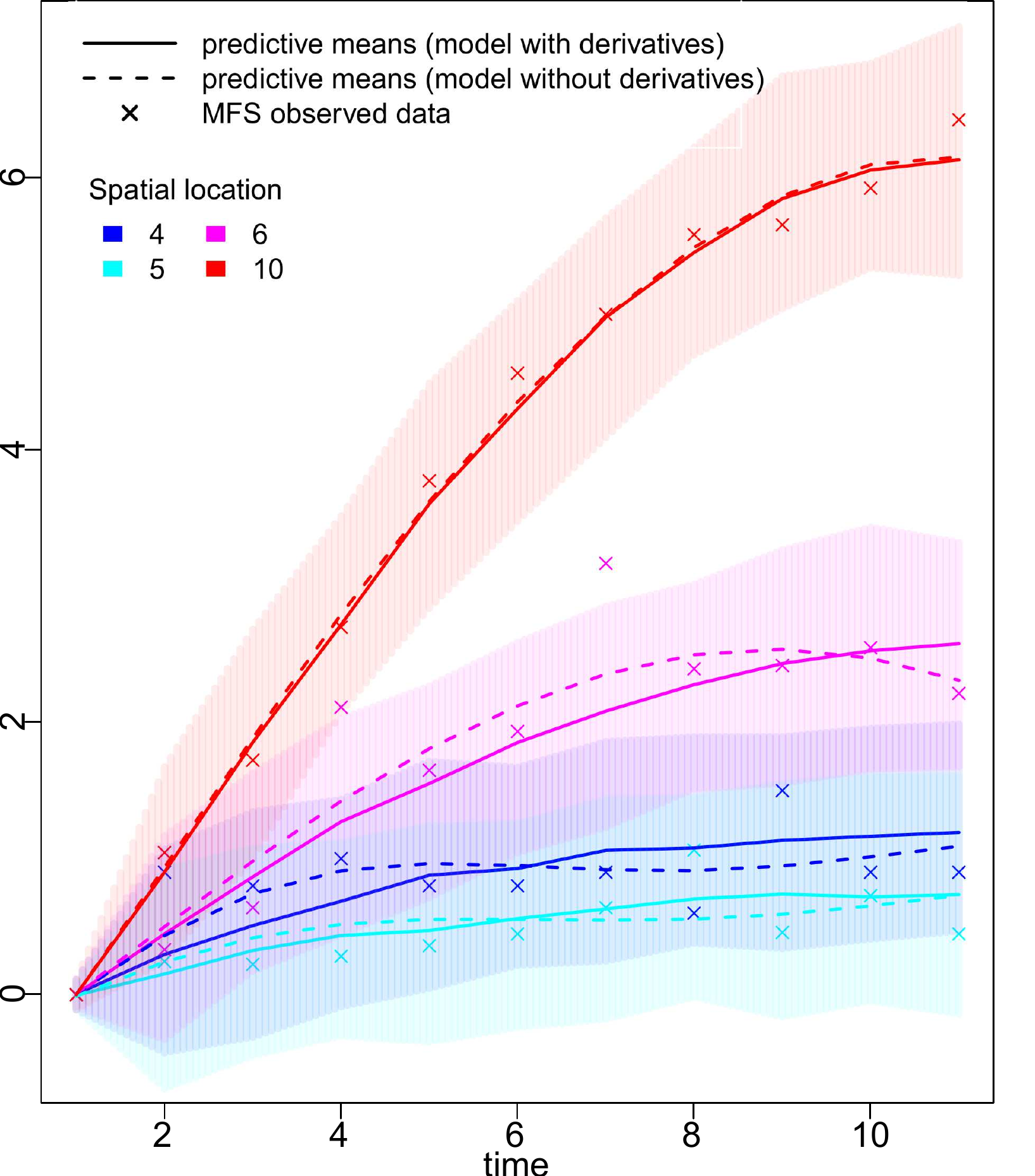}}
\subfigure{\includegraphics[width=0.325\textwidth,trim = 0mm 0mm 165mm -5mm, clip]{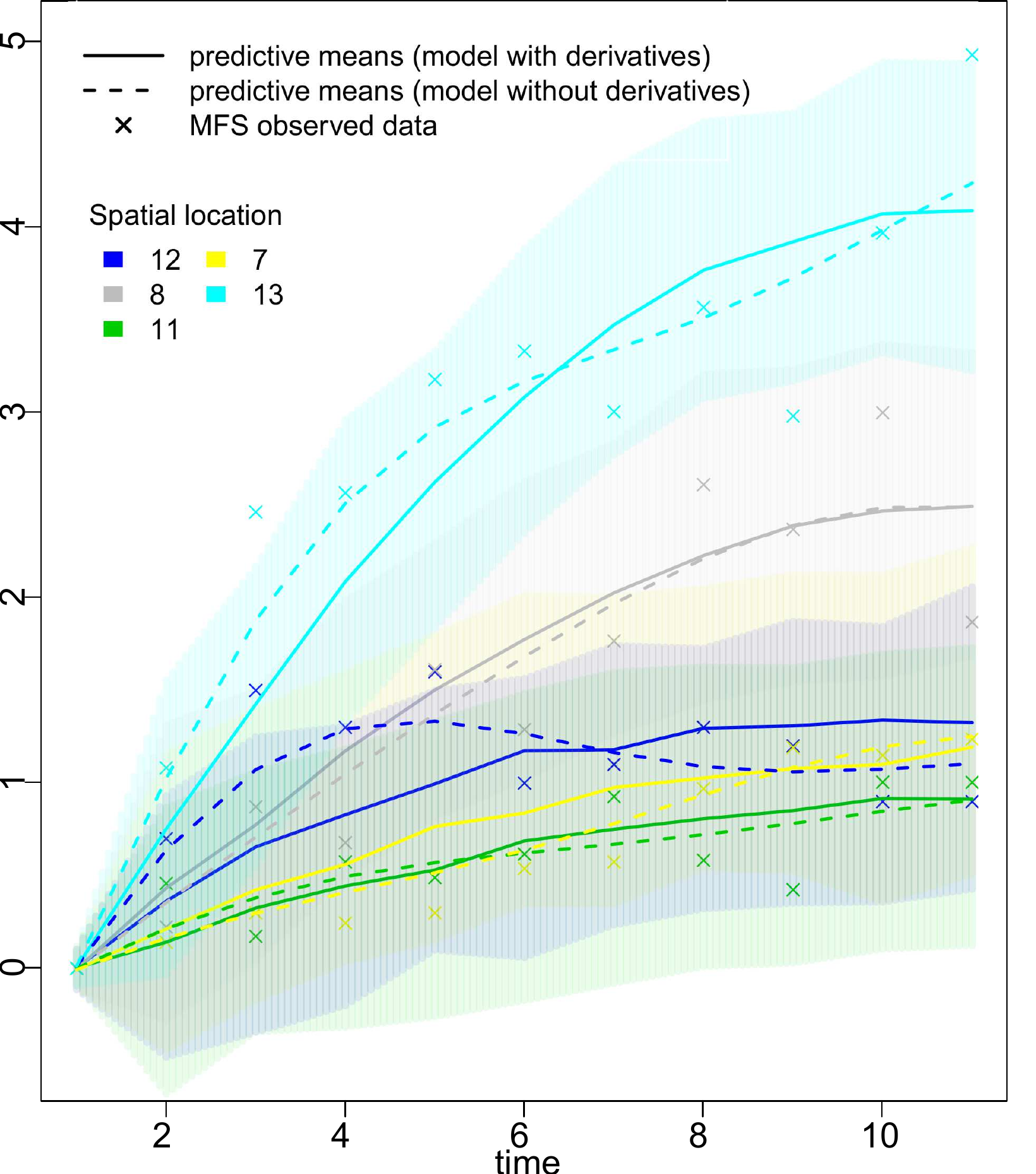}}
\caption{Predictive means $\tilde{\bm{y}}$ and 95\% pointwise credible intervals of the regular process $p(\tilde{\bm{y}}|\bm{y},\bm{y}')$ as a function of time for specific spatial locations using both models with and without derivative information, and superimposed to the actual of MFS time-series. } \label{fig:fitted}
\end{figure*}

\begin{figure*}[htbp]
\centering
\subfigure{\includegraphics[width=1\textwidth]{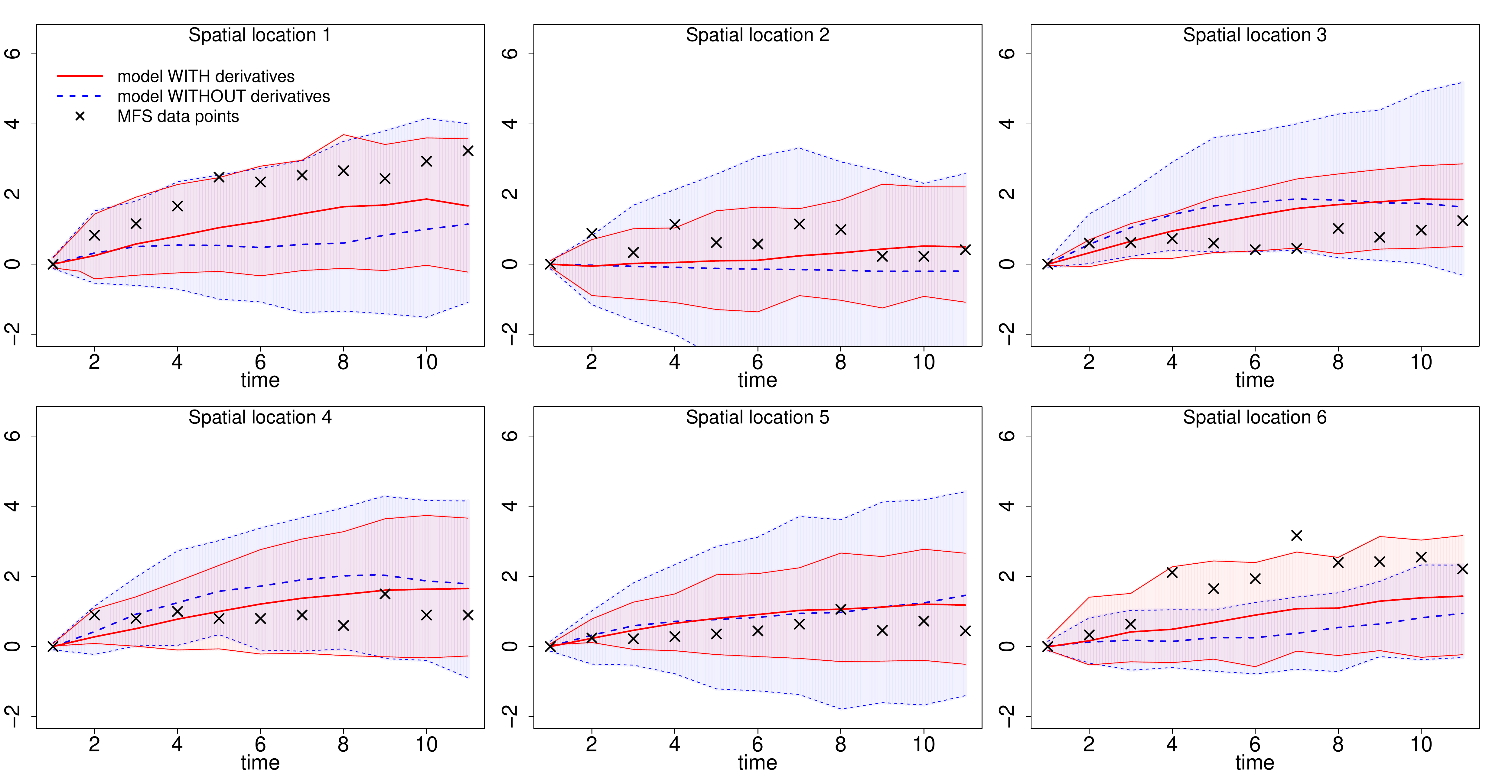}}
\subfigure{\includegraphics[width=1\textwidth,trim = 0mm 0mm 0mm 10mm]{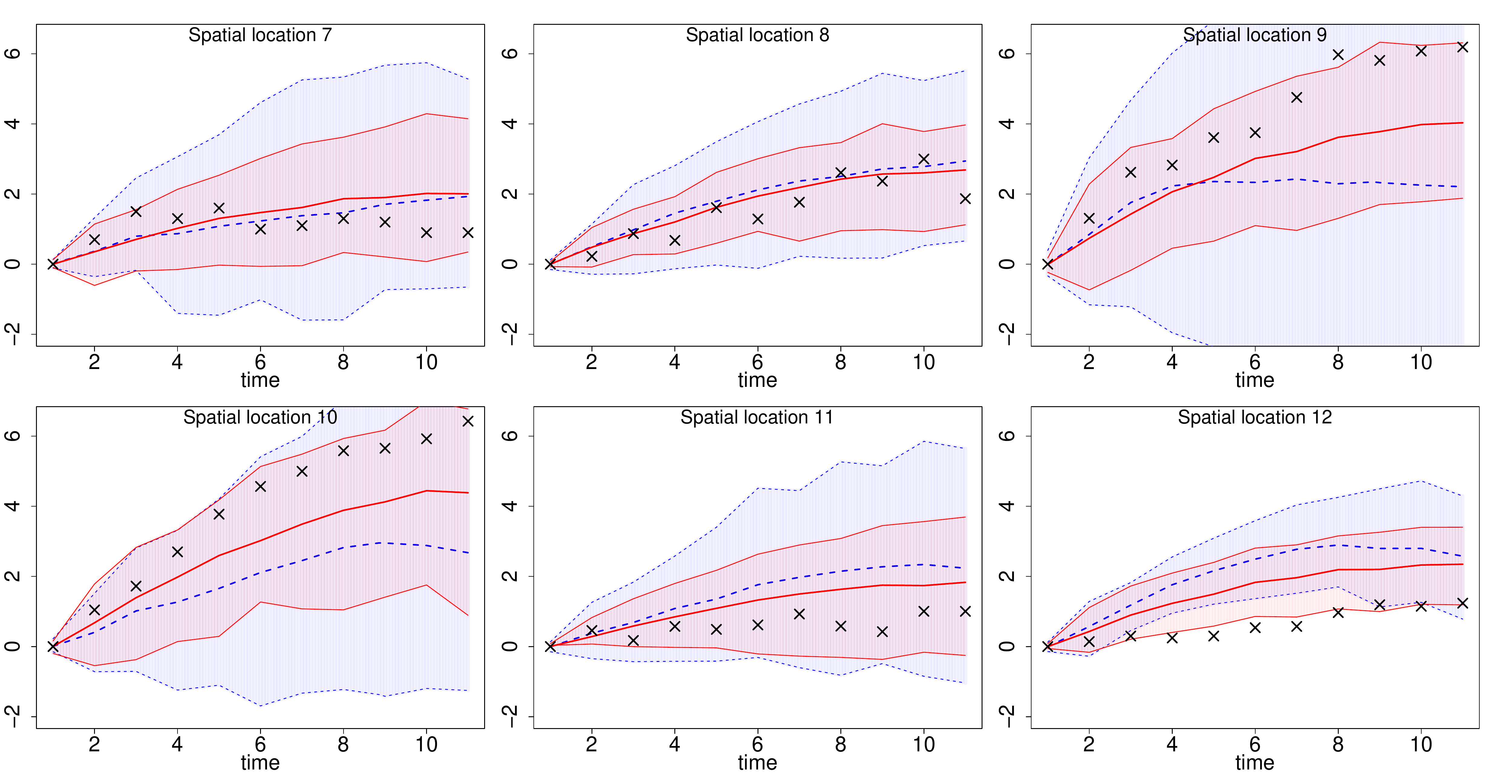}}
\caption{Predicted time-series at new locations in the leave one location out cross-validation procedure (CV2), using both models with and without derivative information. The actual data MFS time-series for every spatial location are plotted as crosses.} \label{fig:cross_validation}
\end{figure*}

Figure \ref{fig:cross_validation} shows predictive distributions $p(\tilde{\bm{y}}_i|\bm{y}_{-i},\bm{y}'_{-i})$ of new time-series following the cross-validation scheme CV2, which is based on leaving the whole time-series observartions of the location $i$ out of the training dataset. In this way, the predictive performance of a new time series at an unobserved location is evaluated. Predictive means and pointwise credible intervals for both the model with derivative information and the model without derivative information are plotted for comparison. The actual data of the time-series at predicted locations are also plotted to visually evaluate the predictions.

Table \ref{tab:ELPD} shows the ELPD and MSE statistics computed by following the three different cross-validation scenarios and for both models, i.e. with and without derivatives.

\begin{table}
\caption{\label{tab:ELPD}The ELPD and MSE for both models, with and without derivative information, computed by the three cross-validation scenarios.}

\centering
\fbox{
\begin{scriptsize}
\begin{tabular}{l|l|l|l}
 & & with derivative &  without derivative\\
 & & information & information \\
\hline
\multirow{2}{*}{CV1} & ELPD & -0.61 & -0.78 \\ 
& MSE & 0.14  & 0.15   \\
\hline
\multirow{2}{*}{CV2} & ELPD &  -11.62 & -29.61 \\ 
& MSE & 2.80  &  4.03  \\
\hline
\multirow{2}{*}{CV3} & ELPD &  -4.49 & -9.64 \\ 
& MSE & 0.87  &  1.02  \\
\end{tabular}
\end{scriptsize}
}
\end{table}

\begin{figure}
\centering
\subfigure{\includegraphics[scale=0.24]{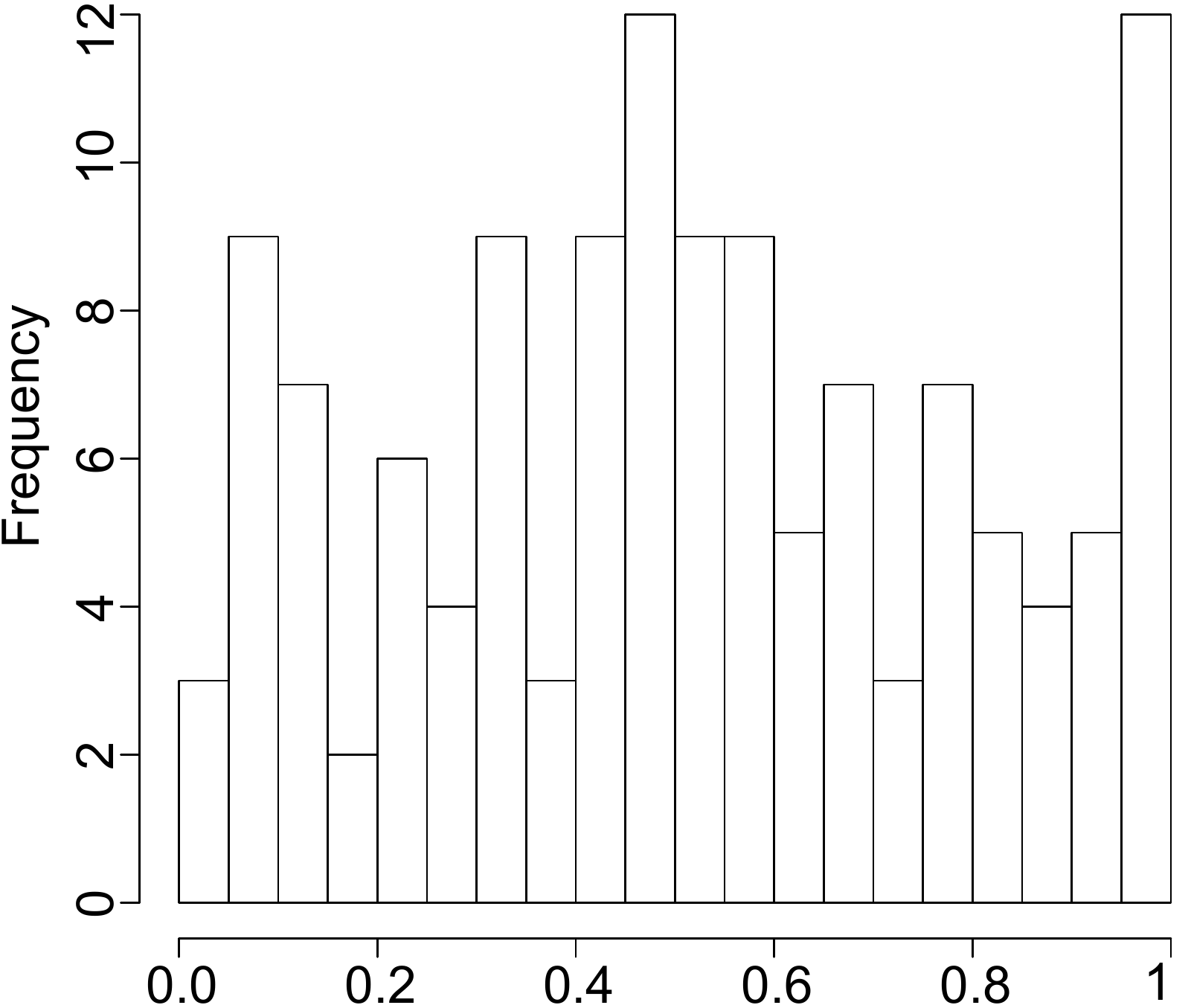}}
\hspace{0.1cm}
\subfigure{\includegraphics[scale=0.24]{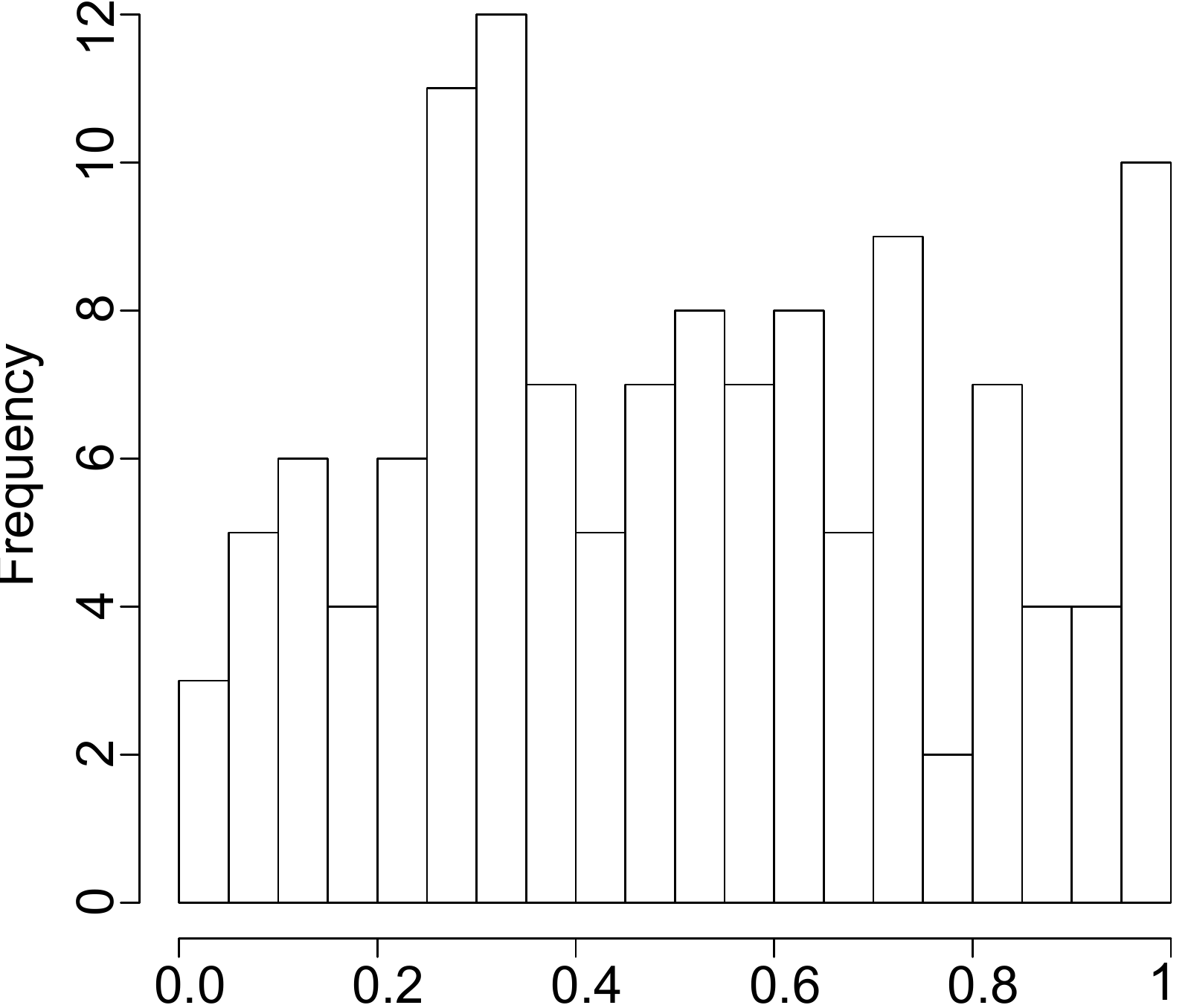}}
\caption{Frequency histograms of the LOO probability integral transformation (LOO-PIT) of both models, with (left) and without (right) derivative information.}
  \label{p_values}
\end{figure}

Figure \ref{p_values} shows the frequency histograms of the LOO probability integral transformation (LOO-PIT) for both models, with and without derivatives.

\section{Discussion \label{discussion}}
The fact that the lengthscale parameters $\rho_1$ and $\rho_3$, corresponding to the variables $H$ and $I$, respectively, are relatively small (the mode of $\rho_1$ is 0.95 and the mode of $\rho_3$ is 1.3) indicates that the function is non-linear with those variables or that the rate of decay of the correlation is high. Therefore, variations in the input variables imply a quick decrease in the correlations allowing for the non-linear effects. 
The variables $S$, $S_x$, and $S_y$ have larger lengthscales (mode of $\rho_2$ is 18.3 and mode of $\rho_4$ is 4.5) so that the function depends on $S$, $S_x$, and $S_y$ in a smoother and less non-linear way. Especially, the variable $S$ with a lengthscale of 18.3 contributes with a constant effect of one to the correlation and implies a constant function to the Gaussian process, so being an irrelevant variable to the process. 

The Frequency histograms of the LOO probability integral transformation (LOO-PIT) (Figure \ref{p_values}) show similarities to uniform distributions in both models , which means good model performances and that the models are compatible with the observed data.

Additional information of the estimated joint covariance matrix $K$ of the process is provided. Images of the spatio-temporal covariance structure of both the regular and derivatives observations, and for both training and predicting data points, are depicted in Appendix \ref{app_cov_matrix}.

Separability between the spatial and temporal input dimensions can be clearly appreciated in Figure \ref{b-vectors} in Appendix \ref{app_post_vs_inputs}, where the values of the process $\bm{f}$ versus the input variables ($H$, $S$ and $I$) on the spatial dimension are independent on the time points, only differing in their overall scale. 

The mode value of the lengthscale parameter $\rho_5$ associated to the time input dimension is 8.2, with a standard deviation of the variable $t$ of approximately 3, allowing smoothing over the temporal dimension as can be seen in Figure \ref{fig:fitted}.

The use of a noise-free pseudo observation model at the starting time points, $\{(i,t):t=0\}$, forces the predictive distributions $p(\tilde{\bm{y}}|\bm{y},\bm{y}')$ to be zero at those starting points (Figures \ref{fig:fitted} and \ref{fig:cross_validation}). 

Derivatives are always non-negative and approaching zero at the end of both the observed and predicted new time series (see Figure \ref{fig:fitted_and_derivatives} in Appendix \ref{app_pred_derv}), which means the curves are always non-decreasing as a function of time and stabilizing in the long term. Inducing monotonicity by means of virtual sign observations of the partial derivatives only in the data points $\{(i,t):t\in \{6,9\}\}$ has been sufficient to achieve monotonicity throughout of the time-series. 


Better predictive performance of new time-series at unobserved locations is appreciated for the model with derivatives (Figure \ref{fig:fitted}). 
Closer predictions to the actual data, narrower predictive intervals and good dynamics of the functions for the model with derivatives can be apprecited. The model without derivatives shows decreasing patterns on the functions which are not consistent with the prior knowledge. 

Monotonicity and long term saturation properties of the curves were not ensured using the model without derivative information (Figures \ref{fig:fitted} and \ref{fig:cross_validation}). Hence, the proposed model with derivative information yields a better fit and predictions for the functions dynamics, improving their interpretability. In this sense, the analysis of the color fading curves using a model without derivative information could not be done properly because the temporal degradation, specially at the last time points, is unrealistic.


The MSE of the model with derivatives is slightly lower than the model without derivatives, especially in the cross-validation scheme CV2 that means lower mean error of predicting new time-series at unobserved spatial locations, and CV3 where the extrapolation is correlated. Furthermore, when the uncertainty is taken into account in the evaluation with the ELPD statistic, the improvement of using derivatives is even considerably larger.

Prediction sensitivity due to the small set of data available have been found. It can be seen in Figure \ref{fig:cross_validation} poor predictive performance in some spatial locations when compared to the observed data. This is due to the high sensitivite of the model to leaving some data due to the data set is small. 

The order of the covariance matrix is of $NT \times NT$, requiring $O(NT^3)$ computation expense in the matrix inversion. This operation needs to be repeated at each MCMC step with changing hyperparameters. This prevents using sampling methods for Bayesian inference on Gaussian process to fit and predict large data set, since the computational expenses increase rapidly with $NT$. In case of large data set, distributional approximation methods are recommended. 

When derivative sign observations are included in the model, the posterior predictive distribution conditioned on the hyperparameters in eq. (\ref{eq_pred_posterior}) is no longer Gaussian. This implies that the joint posterior predictive distribution is no longer available in analytical form (see Appendix \ref{app_analytic_predictive}) and sampling from the predictive distribution or distributional approximations methods must to be used.

\section{Conclusion \label{conclusion}}
Color is an important aspect in documentation and conservation of the historical materials, such as rock art paintings, so the knowledge of the potential color degradation on these systems is crucial for eventual safeguarding and conservation. MFS measurements are difficult and lengthy to materialize, especially in these rock art systems, so an interpolation procedure in order to make predictions for new unobserved locations on the surface is important. Furthermore, these measurements in these systems are contaminated with large fluctuations, so the consideration of constraints in the modelling in order to overcome possible modelling issues that may arise due to these large fluctuations is highly encouraged. 

A procedure for jointly modelling a spatio-temporal Gaussian process and its derivative process has been developed for modelling and predicting the spatio-temporal MFS data. GP is a natural and flexible non-parametric prior model for $D$-dimensional functions and with multivariate predictors in each dimension. A GP model exploit at maximum the covariance estructure by means of its covariance function. Furthermore, the GP has been extended to jointly model the regular and derivatives observations, and estimate a joint covariance function between regular and derivative, making it a more informative and rich model and, at the same time, guaranteeing that the functions are non-decreasing as a function of time. Taking into account this first order constraints demonstrated being beneficial, either in terms of predictive performance or application-specific interpretability. Predictive capacity (MSE ans ELDP) are considerably better with the model with derivatives compared to the model without derivatives. However, high sensitivite of the model to leaving some data has been found due to the data set is small. 


A multivariate covariance function in a GP has allowed the usage of many predictors to evaluate the covariance structure of the data. In this sense, we have been able to include the color space variables, the spatial distance and the time points in the covariance function, and demonstrated the colorimetric variables being useful to correlate MFS data. The contribution of the spatial positions on this covariance structure has been found to be quite weak, consequently, often and widely used traditional spatial or spatio-temporal models are not be able of detecting a useful correlation structure among this type of data. 


To summarize, multivariate covariance metrics and zero and first order constraints might be very hard to implement outside of a Bayesian framework and Gaussian process models. Gaussian Process is flexible enough which allowed us to properly model this complex spatio-temporal phenomema dependent on different input covariates and determinants of the behavior of the functions. 

\begin{appendices}

%
%
%

\section{Marginal distributions of the hyperparameters}\label{app_hist_post}
\begin{figure*}
  \centering
  \includegraphics[width=1\textwidth]{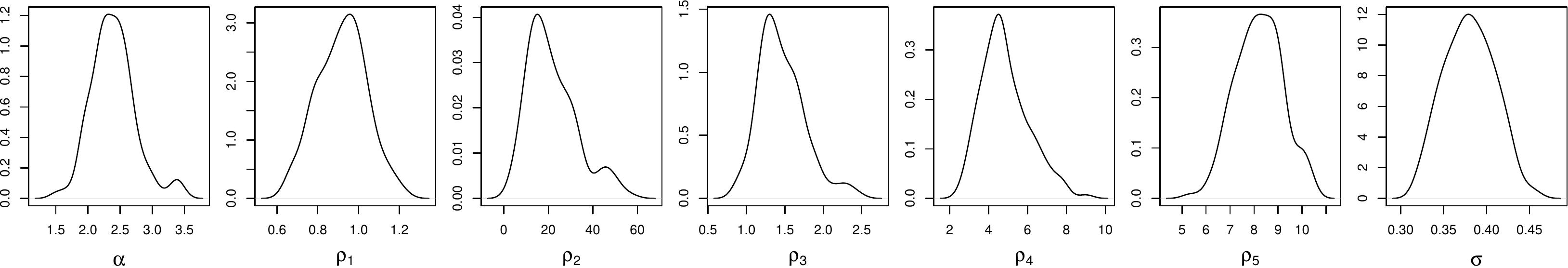}
  \vspace{-0.5cm}
\caption{Marginal posterior density distributions for the hyperparameters $\theta=(\alpha,\bm{\rho}= \{\rho_d \}^{D=5}_{d=1},\sigma)$.}
\label{param}
\end{figure*}

Figure \ref{param} shows the plots of the estimated posterior distributions, $p(\theta|\bm{y},\bm{y}')$, for the hyperparameters $\theta$ of the model. The hyperparameter $\alpha$ is the overall scale of the latent Gaussian process prior. The hyperparameter $\sigma$ is the standard deviation of the observations. The hyperparameter $\rho_d$ ($d=1,\dots,5$) is the lengthscale parameter of the exponentiated quadratic covariance function of the Gaussian process prior associated with the $d$th input variable.

\section{Predictive distributions versus the predictors}\label{app_post_vs_inputs}
The posterior distributions of the process, $p(\bm{f}|\bm{y},\bm{y}')$, versus the input variables $H$, $S$ and $I$, and the time points, are plotted in Figure \ref{b-vectors} in Appendixv\ref{app_post_vs_inputs}. The variables $H$, $S$ and $I$ belong to the spatial dimension. The function $\bm{f}$ is plotted for the different time points.

\begin{figure*}
\centering
  \includegraphics[width=1\textwidth]{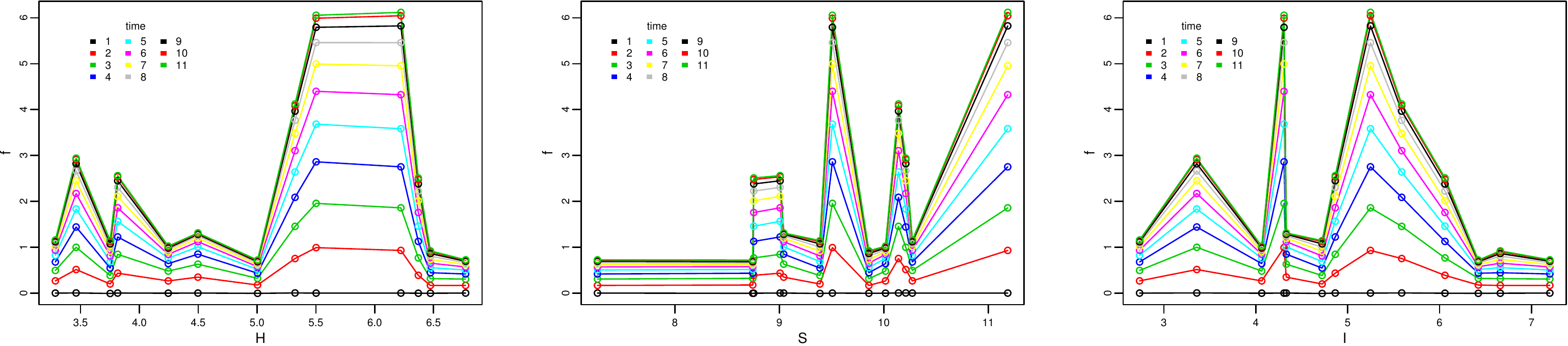} 
  \vspace{-0.5cm}
  \caption{Posterior means of the latent Gaussian process $\bm{f}$ versus the input variables $H$, $S$, and $I$, for specific time points.}
  \label{b-vectors}
\end{figure*}

\section{Predictive distributions of the derivative process}\label{app_pred_derv}
\begin{figure*}[htbp]
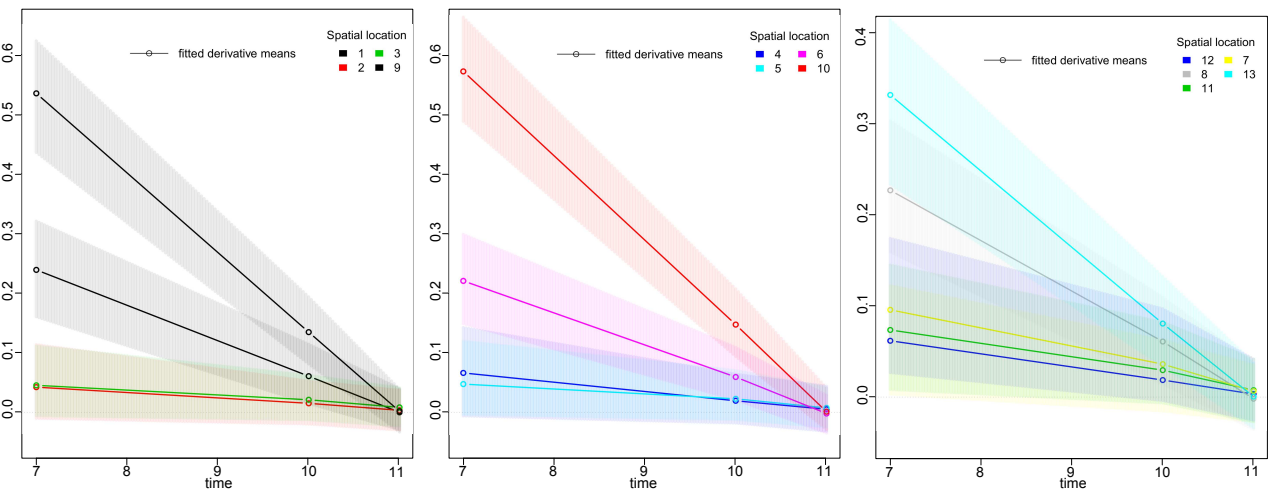

\centering
\subfigure{\includegraphics[width=0.325\textwidth,trim = 165mm 0mm 0mm 0mm, clip]{fitted_and_derivatives_1.pdf}}
\subfigure{\includegraphics[width=0.325\textwidth,trim = 165mm 0mm 0mm 0mm, clip]{fitted_and_derivatives_2.pdf}}
\subfigure{\includegraphics[width=0.325\textwidth,trim = 165mm 0mm 0mm 0mm, clip]{fitted_and_derivatives_3.pdf}}
\caption{ Posterior derivatives means $\bm{f}'$ and 95\% pointwise credible intervals of the derivative process $p(\bm{f}'|\bm{y},\bm{y}')$ as a function of time for specific locations. } \label{fig:fitted_and_derivatives}
\end{figure*}

Figure \ref{fig:fitted_and_derivatives} shows the predictive distributions (predictive means and 95\% pointwise predictive credible intervals) of the derivative process, $p(\bm{f}'|\bm{y},\bm{y}')$, as a function of time for specific spatial locations. The derivative process has only been observed at a subset of time points: values of the sign of the partial derivatives, where we induce monotonicity, have been observed at the time points $\{(i,t):t\in\{6,9\}\}$.

\section{Posterior covariance matrix}\label{app_cov_matrix}
\begin{figure*}[htbp]
\centering
  \subfigure{\includegraphics[scale=0.85,trim = 0mm 18mm 0mm 0mm, clip]{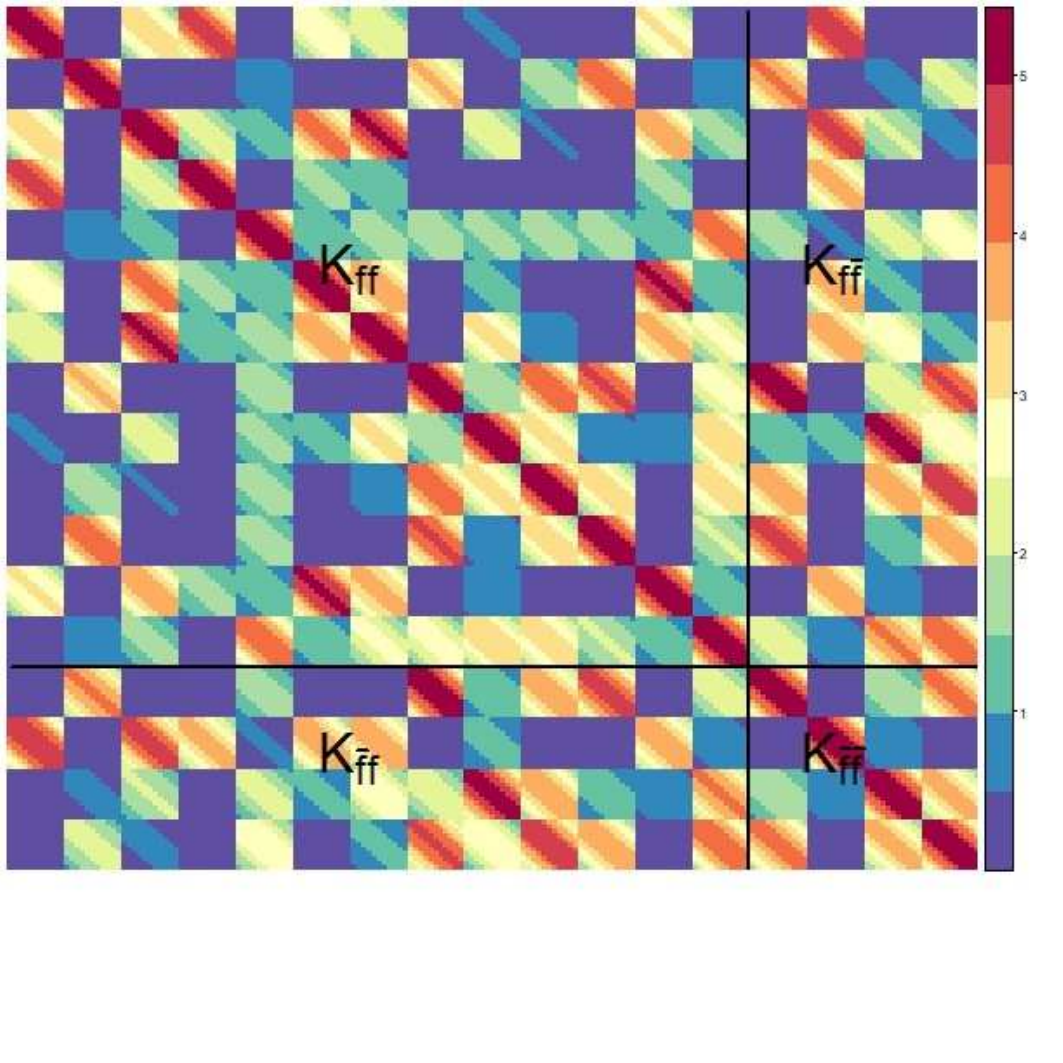}}
  \hspace{0.5cm}
  \subfigure{\includegraphics[scale=0.30]{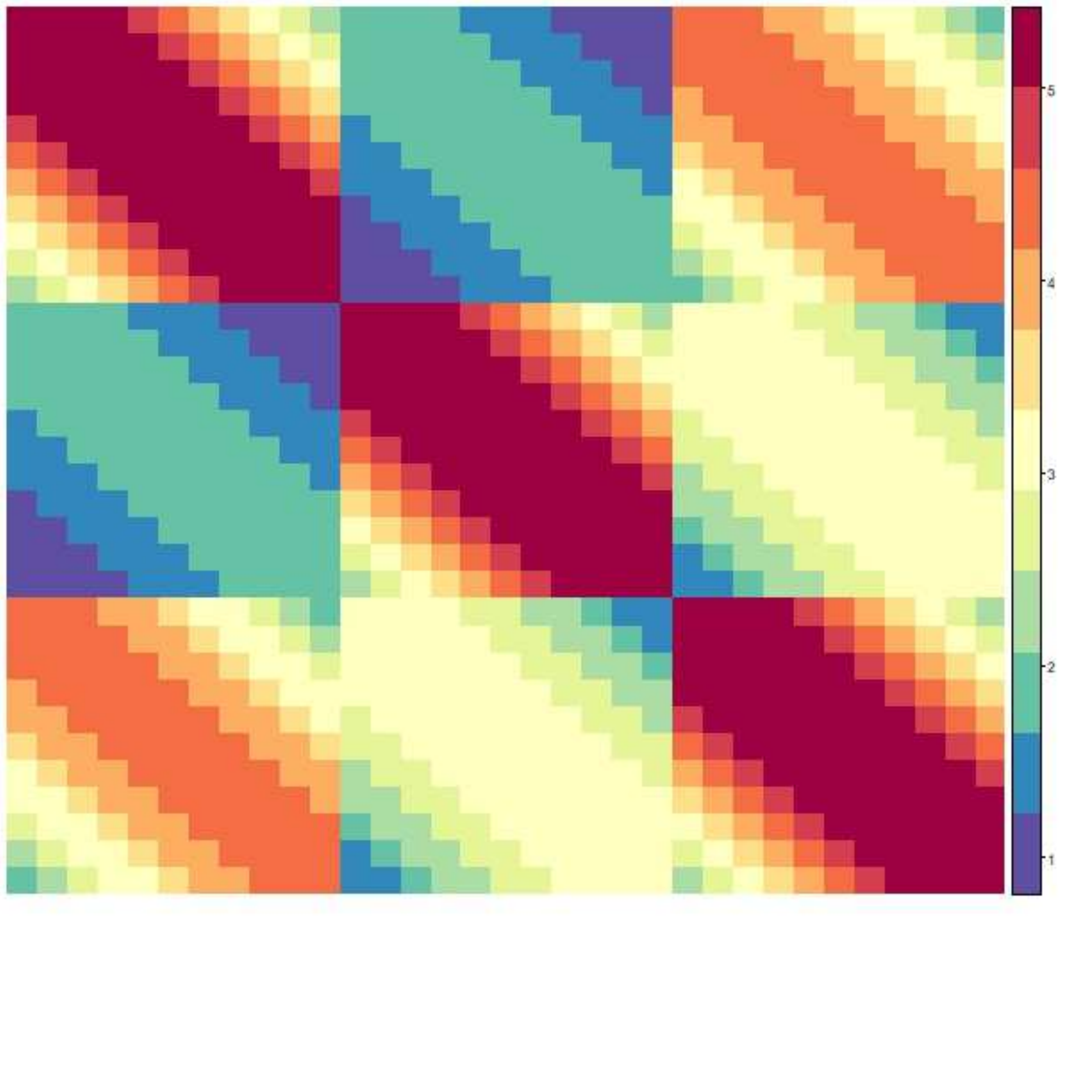}}
  \caption{(left) Covariance matrix image for the posterior distribution $p(\bm{f}|\bm{y},\bm{y}')$ and predictive distribution $p(\bm{f}'|\bm{y},\bm{y}')$ of the regular process. The subindexes of the matrix $K$ denote the type of observations that are involved in each covariance block: $f$ - training data points; $\tilde{f}$ - predicting data points. (right) Extract from the covariance matrix containing the spatio-temporal covariances of the time-series of three locations.} \label{covmat1}
\end{figure*}

\begin{figure*}
\centering
  \includegraphics[width=0.7\textwidth]{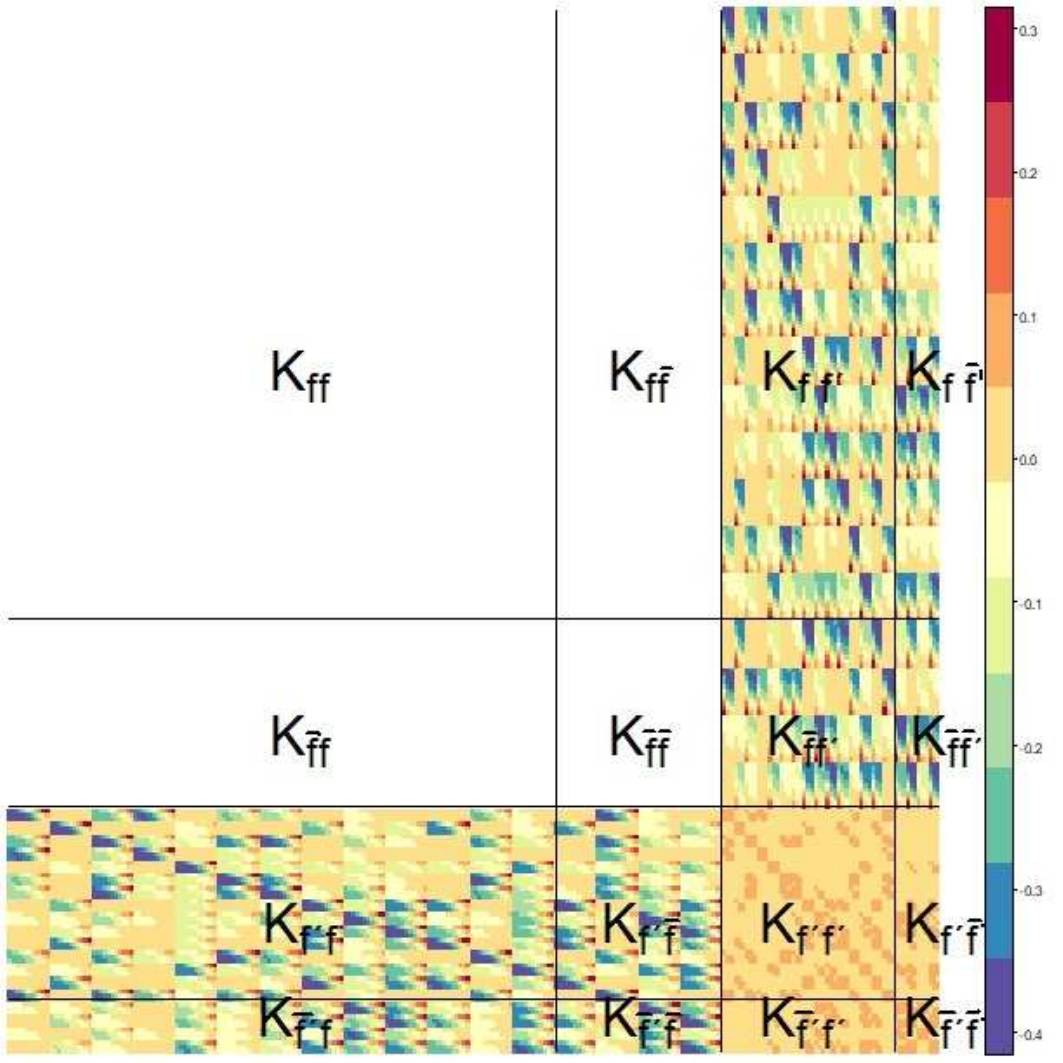}
  \vspace{-0.0cm}
  \caption{Covariance matrix image between the process and its derivatives. The subindexes of the matrix $K$ denote the type of observations that are involved in each covariance block: $f$ - regular observations of training points; $\tilde{f}$ - regular observations of predicting points; $f'$ - derivative observations of training points;  $\tilde{f}'$ - derivative observations of predicting points.}
  \label{covmat2}
\end{figure*}

The joint covariance matrix $K$ of the process is visualized in Figures \ref{covmat1} and \ref{covmat2}. 
Figure \ref{covmat1}(left) shows the part of the covariance matrix that involves the regular process. Figure \ref{covmat1}(right) shows a submatrix of the covariance matrix of the regular process, in which the spatio-temporal covariance structure of three spatial locations and their time-series can be appreciated. The black lines in Figure \ref{covmat1}(left) divide the covariance matrix according to whether it involves covariances among the 13 training data ($K_{ff}$) or among the 4 predicting data ($K_{\tilde{f} \tilde{f}}$) or among the interaction between training and predicting data ($K_{f \tilde{f}}$ and $K_{\tilde{f} f}$). 

Figure \ref{covmat2} shows the parts of the covariance matrix that involve the regular process and its derivatives (derivative process). The block $K_{ff'}$ contains the covariances among regular and derivative observations for the training data. The block $K_{f'f'}$ contains the covariances among derivative observations for the training data. The block $K_{\tilde{f}\tilde{f}'}$ contains the covariances among regular and derivative observations for the predicting data. The block $K_{f' \tilde{f}'}$ contains the covariances among derivative observations for training and predicting data. And, finally, the block $K_{\tilde{f}' \tilde{f}'}$ contains the covariances among derivative observations for predicting data.

\section{Analytical form of predictive distribution of a GP}\label{app_analytic_predictive}
If the likelihood functions for both regular and derivative observations were Gaussian, Bayesian predictive inference can be sped up by using the analytical form of the posterior predictive distribution, and then sampling directly from it. This posterior predictive distribution can be derived from the joint Gaussian distribution for observations $y_{\mathrm joint}$ and mean predictions $\mu_{\mathrm joint}$ for new inputs $\tilde{X}$. Thus, the conditional distribution of predictions given the data and parameters is obtained from the properties of this multivariate Gaussian (\ref{eq:analyticprediction}).
\begin{figure*}[!t]
\begin{align} \label{eq:analyticprediction}
& p(\bm{\tilde{\mu}}_{\mathrm joint}|\bm{y}_{\mathrm joint},X,\tilde{X},\theta,\sigma^2) = \mathcal{N}(\bm{\tilde{\mu}}_{\mathrm joint}|\mathrm{E}(\bm{\tilde{\mu}}_{\mathrm joint}),\mathrm{cov}(\bm{\tilde{\mu}}_{\mathrm joint})) \nonumber  \\
& \mathrm{E}(\bm{\tilde{\mu}}_{\mathrm joint}) = K(\tilde{X},X,\theta)(K(X,X,\theta)+\sigma^2I)^{-1}\bm{y}  \\
& \mathrm{cov}(\bm{\tilde{\mu}}_{\mathrm joint}) = K(\tilde{X},\tilde{X},\theta)-K(\tilde{X},X,\theta)(K(X,X,\theta)+\sigma^2I)^{-1}K(X,\tilde{X},\theta). \nonumber
\end{align}
\end{figure*}

\end{appendices}

\section*{Acknowledgements}

The authors want to thank The National Museum in Krakow and Dr. Julio M. del Hoyo-Mel\'endez for providing the MFS instrumentation to collect the data. The authors gratefully acknowledge the support from the Spanish Ministerio de Econom\'ia y Competitividad to the project HAR2014-59873-R, and the Academy of Finland Grant 298742.

\bibliographystyle{biom} \bibliography{references}

\label{lastpage}

\end{document}